\documentclass[twocolumn,aps,prl,amsmath,amssymb,preprintnumbers]{revtex4}
\begin{document}

\title{Comment (2) on ``Quantum Transfiguration of Kruskal Black Holes''}

\author{Martin Bojowald}
\email{bojowald@gravity.psu.edu}
\affiliation{Institute for Gravitation and the Cosmos,
The Pennsylvania State University, 104 Davey Lab, University Park,
PA 16802, USA}

\begin{abstract}
 An explicit calculation shows that the model constructed in \cite{Transfig}
violates general covariance. It is therefore unphysical and does not determine
properties of space-time or black holes.
\end{abstract}

\maketitle

The authors of \cite{Transfig} attempt to modify the dynamics of gravitational
collapse by viewing the static, spherically symmetric exterior space-time
region of a Kruskal black hole as a homogeneous model with timelike
hypersurfaces ``evolving'' in the direction of the radial coordinate.
However, the authors never consider the spherically symmetric, static
description after they modify the dynamics.  This description, in fact, is no
longer equivalent, owing to a breakdown of general covariance in the model.

An $\hbar$-dependent effective theory is covariant if and only if it is
covariant order by order in $\hbar$. It is therefore sufficient to consider
only leading-order modifications, parameterized in \cite{Transfig} by
$\delta_b$ and $\delta_c$. This leads to the cosmological Hamiltonian
\begin{eqnarray} \label{Hhom}
 H_{\rm hom}[N]&=&-\frac{N}{2G}
 \left(2bc\sqrt{p_c}+(1-b^2)\frac{p_b}{\sqrt{p_c}}\right.\\
&&\left.+ 
\frac{1}{3}b^3(c\sqrt{p_c}-bp_b/\sqrt{p_c})\delta_b^2
- \frac{1}{3}bc^3\sqrt{p_c}\delta_c^2\right)\nonumber
\end{eqnarray}
for an arbitrary lapse function $N$, Newton's constant $G$, and canonical
pairs $(p_b,b)$ and $(c,p_c)$.  (In the specific choice of \cite{Transfig},
$\delta_c$ is of higher order in $\hbar$ than $\delta_b$ and could therefore
be eliminated here without changing our result. We assume $p_c>0$ and ignore
parameters such as $\gamma$ and $L_0$ that are irrelevant for our argument.)

The meaning of $p_b$ and $p_c$ is shown by how they appear in the space-time
line element
\begin{equation} \label{ds1}
 {\rm d}s^2= -\frac{p_b^2}{p_c}{\rm d}x^2+N^2{\rm d}t^2+ p_c{\rm
   d}\Omega^2\,,
\end{equation}
while $b$ and $c$ are related to time derivatives of $p_b$ and $p_c$ according
to equations of motion generated by (\ref{Hhom}). Inverting these equations,
we obtain
\begin{eqnarray} \label{bc}
 b&=& \frac{\dot{p}_c}{2N\sqrt{p_c}}- \frac{1}{6}b^3\delta_b^2+
 \frac{1}{2}bc^2 \delta_c^2\\
c&=& -\frac{\dot{p}_b}{N\sqrt{p_c}}+ b\frac{p_b}{p_c} -
\frac{1}{6}b^2(3c-4bp_b/p_c)\delta_b^2+ \frac{1}{6} c^3\delta_c^2\,. \label{bcc}
\end{eqnarray}
They are valid if $\delta_b$ and $\delta_c$ depend on $p_b$ or $p_c$.  If
$\delta_b$ and $\delta_c$ also depend on $b$ or $c$, (\ref{bc}) and
(\ref{bcc}) acquire extra terms that do not change our result of
non-covariance. We work with (\ref{bc}), (\ref{bcc}) because their solutions
(to all orders in $\delta_b$ and $\delta_c$) have been used in
\cite{Transfig}; see also \cite{DiracPoly}.

We now evaluate the same dynamics from the perspective of a spherically
symmetric model with line element
\begin{equation} \label{ds2}
 {\rm d}s^2= -M^2{\rm d}T^2+\frac{(E^{\varphi})^2}{E^x}{\rm d}X^2+ E^x{\rm
   d}\Omega^2
\end{equation}
where $E^x$ and $E^{\varphi}$ are the components of a densitized triad
analogous to $p_c$ and $p_b$. Comparing (\ref{ds1}) and (\ref{ds2}) such that
$T=x$ and $X=t$, we obtain the correspondence $p_c=E^x$, $p_b=M\sqrt{E^x}$,
$N=E^{\varphi}/\sqrt{E^x}$, while $b$ and $c$ follow from (\ref{bc}) and
(\ref{bcc}) observing that the dot amounts to an $X$-derivative
(a prime) in spherical symmetry.

{\em If} there is a covariant space-time underlying this model, the
Hamiltonian constraint (\ref{Hhom}), using our correspondence, must vanish
when expressed in spherically symmetric models.  After several concellations,
we obtain
\begin{eqnarray} \label{Hhomsph}
 -2GH_{\rm hom}[N]=-\frac{M'(E^x)'\sqrt{E^x}}{E^{\varphi}}+
   \frac{ME^{\varphi}}{\sqrt{E^x}}-
   \frac{M((E^x)')^2}{4\sqrt{E^x}E^{\varphi}}\\
+
   \frac{1}{8}\frac{(M\sqrt{E^x})'((E^x)')^3}{(E^{\varphi})^3} \delta_b^2-
   \frac{1}{2}
   \frac{(M')^3(E^x)'(E^x)^{3/2}}{(E^{\varphi})^3}\delta_c^2\,.\nonumber
\end{eqnarray}

The lapse function $M$ of the spherically symmetric model is not independent
but rather determined (non-locally) by $E^x$ and $E^{\varphi}$ through
staticity (zero momenta). This condition amounts to a differential equation
\begin{equation}
M'+\frac{1}{4}\frac{(E^x)'}{E^x}M+\frac{1}{2}
\frac{(E^{\varphi})^2}{\sqrt{E^x}(E^x)'}V(E^x)M=0\,, 
\end{equation}
which, for full generality, we use here for a general dilaton model
\cite{Strobl} with potential $V(E^x)$; see for instance
\cite{SphSymmPSM,MidiClass}.  (These models are broad and include, for
instance, spherically symmetric Palatini-$f(R)$ theories through their
equivalence with scalar-tensor metric theories with a non-dynamical scalar
\cite{PalatinifR}.)  Classically, (\ref{Hhomsph}) implies
\begin{equation}
 -2GH_{\rm hom}[N]= \frac{ME^{\varphi}}{\sqrt{E^x}}
 \left(1+\frac{1}{2}\sqrt{E^x}V(E^x)\right)=0
\end{equation}
which selects the correct dilaton potential $V(E^x)=-2/\sqrt{E^x}$ of
classical spherically symmetric gravity.  Non-zero $\delta_b$ and $\delta_c$
imply additional terms in (\ref{Hhomsph}) which depend on $M$ and
$E^{\varphi}$. They cannot be absorbed in a modified dilaton potential
$V(E^x)$.  Since dilaton gravity actions present the most general class of
generally covariant $1+1$-dimensional theories, the model of \cite{Transfig}
violates covariance.

{\em Note added:} A transformation of the cosmological-type solutions given in
\cite{Transfig} to spherically symmetric slices has already been performed in
\cite{TransCommAs}, focusing on implications for the asymptotic structure. In
the present comment, an extension of such a transformation to the model's
constraints and equations of motion makes it possible to draw conclusions
about covariance.

\medskip

\noindent The author thanks Norbert Bodendorfer and Suddhasattwa Brahma for
comments. This work was supported in part by NSF grant PHY-1607414.


\begin{thebibliography}{1}

\bibitem{Transfig}
A. Ashtekar, J. Olmedo, and P. Singh, Phys.\ Rev.\ Lett. {\bf 121},  241301
  (2018).

\bibitem{DiracPoly}
N. Bodendorfer and J. Mele, F.~M.\ abd~M\"unch,   arXiv:1902.04032.

\bibitem{Strobl}
T. Strobl,   hep-th/0011240.

\bibitem{SphSymmPSM}
M. Bojowald and J.~D. Reyes, Class.\ Quantum Grav. {\bf 26},  035018  (2009).

\bibitem{MidiClass}
M. Bojowald and S. Brahma, Phys.\ Rev.\ D {\bf 95},  124014  (2017).

\bibitem{PalatinifR}
G.~J. Olmo, Phys.\ Rev.\ Lett. {\bf 95},  261102  (2005).

\bibitem{TransCommAs}
M. Bouhmadi-L\'{o}pez {\it et~al.},   arXiv:1902.07874.

\end{thebibliography}
\end{document}